\documentclass[12pt]{article}
\usepackage{latexsym}
\usepackage{axodraw}
\usepackage{epsfig}
\usepackage{amsfonts}
\usepackage{array,longtable}
\begin{document}

 \title{\bf Renormalizable Expansion for Nonrenormalizable Theories:\\
 I. Scalar Higher Dimensional Theories}
\author{D.I.Kazakov}
\date{}
\maketitle \vspace{-0.8cm}
\begin{center}
Bogoliubov Laboratory of Theoretical Physics, Joint
Institute for Nuclear Research, Dubna, Russia \\
Institute for Theoretical and Experimental Physics, Moscow, Russia\\
e-mail: KazakovD@theor.jinr.ru \\[0.5cm]

{\large and G.S.Vartanov}

\vspace{0.8cm}
Museo Storico della Fisica e Centro Studi e Ricerche "Enrico Fermi",
Rome, Italy\\
Bogoliubov Laboratory of Theoretical Physics, Joint
Institute for Nuclear Research, Dubna, Russia \\[0.1cm]
e-mail: Vartanov@theor.jinr.ru\\[0.1cm]
\end{center}

\begin{abstract}
We demonstrate how one can construct renormalizable perturbative expansion in formally
nonrenormalizable higher dimensional scalar theories. It is based on $1/N$-expansion
and results in a logarithmically divergent perturbation theory in arbitrary high odd
space-time dimension. The resulting effective coupling is dimensionless and is running
in accordance with the usual RG equations. The corresponding beta function is calculated in
the leading order and is nonpolynomial in effective coupling. It exhibits either UV
asymptotically free or IR free behaviour depending on the dimension of space-time.
\end{abstract}

\section{Introduction}

Popular nowadays  higher dimensional theories~\cite{Multi} suffer from the lack of
renormalizable perturbative expansion. The usual coupling has a negative dimension, thus
leading to power increasing divergencies which are out of control. Popular reasoning when
dealing with such theories relies on higher energy (string) theory which is supposed to
cure all the UV problems while the low energy one is treated as an effective theory
basically at the tree level.

In our previous work~\cite{our} we considered the scalar theories
in extra dimensions within the usual perturbative expansion and
demonstrated that the leading divergences are governed by the
one-loop diagrams even in the nonrenormalizable case, as it was
shown in~\cite{Kazakov}. Contrary to that work, we make here an
attempt to construct renormalizable expansion in such formally
nonrenormalizable theories. As in~\cite{our}, we consider scalar
higher dimensional theories as an example, but here we treat them
in the framework of $1/N$-expansion(for review see~\cite{Justen}).
The resulting perturbation theory is shown to be renormalizable,
logarithmically divergent in any odd dimension $D$ and obtains an
effective dimensionless expansion parameter.

We show below how one can construct such expansion and calculate the leading terms.
One finds that resulting PT is nonpolynomial in effective coupling, but polynomial in
$1/N$ and obeys the usual properties of renormalizable theory. It might be either UV
asymptotically free or IR free depending on the space-time dimension $D$.

\section{$1/N$-expansion}

Let us start with the usual $N$ component scalar field theory in $D$ dimensions, where
$D$ takes an arbitrary odd value, with $\phi^4$ self-interaction. The Lagrangian looks like
\begin{equation}\label{l}
  {\cal L} = \frac 12 (\partial_\mu \vec{\phi})^2-\frac12 m^2 \vec{\phi}^2 -
  \frac{\lambda}{8N}(\vec{\phi}^2)^2,
\end{equation}
where $N$ is the number of components of $\phi$. It is useful to rewrite it introducing a
Lagrange multiplier $\sigma$~\cite{Arefeva}
\begin{equation}\label{s}
 {\cal L} = \frac 12 (\partial_\mu \vec{\phi})^2-\frac12 m^2 \vec{\phi}^2 -
 \frac{\sqrt{\lambda}}{2\sqrt{N}}\sigma(\vec{\phi}^2)+\frac 12 \sigma^2.
\end{equation}
Now one has two fields, one $N$ component and one singlet with triple interaction. Let us
look at the propagator of the $\sigma$ field. At  the tree level it is just "i", but then
one has to take into account the corrections due to the loops of $\phi$ (see Fig.1).
\begin{center}
\begin{picture}(300,100)(50,-70)
\SetWidth{1.5}
\DashLine(0,0)(45,0){3}
\SetWidth{0.5}
\Text(52,0)[]{=}
\DashLine(60,0)(90,0){3}
\Text(98,0)[]{+}
\DashLine(105,0)(135,0){3}
\Oval(150,0)(15,15)(360)
\DashLine(165,0)(195,0){3}
\Text(203,0)[]{+}
\DashLine(210,0)(240,0){3}
\Oval(255,0)(15,15)(360)
\DashLine(270,0)(300,0){3}
\Oval(315,0)(15,15)(360)
\DashLine(330,0)(360,0){3}
\Text(373,0)[]{+...}
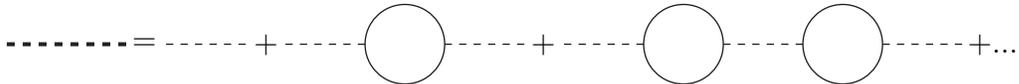
 \Text(186,-30)[]{Figure 1: The chain of diagrams giving a contribution to the
 $\sigma$ field propagator
 in}
 \Text(63,-48)[]{the zeroth order of $1/N$ expansion}
 \label{prop}
\end{picture}
\end{center}\vspace{-0.3cm}

If one follows the $N$ dependence of the corresponding graphs, one finds out that it
cancels: they are all of the zeroth order in $1/N$. Thus, one can sum them up and get
\begin{equation}\label{ppp}
  \textbf{- - - -} = \textmd{- - - -} (\frac{1}{1- O \textmd{- - -}})
  =\frac{i}{1+\lambda f(D)(-p^2)^{D/2-2}},
\end{equation}
where $$f(D)=\frac{\Gamma^2(D/2-1)\Gamma(2-D/2)}{2^{D+1}\Gamma(D-2)\pi^{D/2}}$$ and we
put $m=0$ for simplicity.

Notice that since $\lambda$ is positive, when $f(D)>0$ the obtained propagator has no
pole in the Euclidean region and has a cut for $p^2>0$. If $m\neq 0$, the cut starts at
the threshold $p^2=4m^2$. On the contrary, when $f(D)<0$ one has a pole in the Euclidean
region which may cause trouble when integrating. Due to the factor $\Gamma(2-D/2)$,
$f(D)$ may have any sign depending on the value of $D$.  In the case of negative $f(D)$
we take the integrals in a sense of a principle value.  This is similar to what happens
in logarithmic theories with the so-called renormalon chains~\cite{renormalon}, but
contrary to $\log(p^2)$ the power of $p^2$ is positive in the whole Euclidean region.
Notice also that $f(D)$ is finite for any odd $D$ despite naive power counting. This is
due to the use of dimensional regularization: the one-loop diagrams in odd dimensions are
finite since the gamma function has poles only at integer negative arguments and not at
half-integer ones. This phenomenon can also be understood in  other regularization
techniques, but we do not discuss it here.

Thus, we have now the modified Feynman rules: the $\phi$ propagator is the usual one
while the $\sigma$ propagator is given by eq.(\ref{ppp}).  One can now construct the
diagrams using these propagators and the triple vertex having in mind that any closed
cycle of $\phi$ gives an additional factor of $N$ and any vertex gives $1/\sqrt{N}$.

Let us first analyse the degree of divergence. Let us start with the $\phi$ propagator.
If the diagram with two external $\phi$ lines contains $L$ loops, this means that it has
$2L$ vertices, $2L-1$ $\phi$ lines and $L$ $\sigma$ lines. Notice that each $\sigma$ line
now behaves like $1/(p^2)^{D/2-2}$. Then, the degree of divergence is
\begin{equation}\label{w}
  \omega(G)=LD - (2L-1)2-L(D-4)= 2 !
\end{equation}
for any D. Since this is a propagator, the divergence is proportional to $p^2$ and thus
is reduced to the logarithmic one.

Let us now take the triple vertex. If it has $L$ loops, then one has $2L+1$ vertices,
$2L$ $\phi$ lines and $L$ $\sigma$ lines. Then, the degree of divergence is
\begin{equation}\label{w2}
  \omega(G)=LD - (2L)2-L(D-4)= 0 !
\end{equation}
for any $D$. Thus, we again have only logarithmic divergence.

At last, consider the $\sigma$ propagator. In $L$ loops it has $2L$ vertices, $2L$ $\phi$
lines and $L-1$ $\sigma$ lines. The degree of divergence is
\begin{equation}\label{w3}
  \omega(G)=LD - (2L)2-(L-1)(D-4)= D-4.
\end{equation}
This means that in odd $D$ it has no global divergence (again we explore the properties
of dimensional regularization) and the only possible divergencies are those of the
subgraphs eliminated by renormalization of $\phi$ and the coupling. To see this, consider
a genuine diagram for the $\sigma$-field propagator which is shown in Fig.2, where the
blobs denote the 1PI vertex or propagator and vertex subgraphs.
\begin{center}
\begin{picture}(100,110)(0,-60)
%\Text(-10,1)[]{$R'$}
\SetWidth{1.5} \DashLine(0,0)(15,0){3} \DashLine(80,0)(95,0){3} \SetWidth{0.5}
\Vertex(20,0){5} \Vertex(50,25){5} \Vertex(50,-25){5}
%\CCirc(20,0){7}{Black}{Black}
%\CCirc(50,25){5}{Black}{Black}
%\CCirc(50,-25){5}{Black}{Black}
\Oval(50,0)(25,30)(0)
%\Text(100,1)[]{=}
%\SetWidth{1.5}
%\DashLine(105,0)(120,0){3}
%\DashLine(185,0)(200,0){3}
%\SetWidth{0.5}
%\CCirc(125,0){7}{Black}{Black}
%\CCirc(155,25){5}{Black}{Black}
%\CCirc(155,-25){5}{Black}{Black}
%\Oval(155,0)(25,30)(0)
%\Text(205,-2)[]{-}
%\DashCArc(240,0)(30,0,360){3}
%\DashCArc(300,0)(25,0,360){3}
%\DashCArc(355,0)(25,0,360){3}
%\SetWidth{1.5}
%\DashLine(240,25)(240,0){3}
%\DashLine(390,0)(405,0){3}
%\DashLine(435,0)(450,0){3}
%\SetWidth{0.5}
%\Line(240,0)(260,-15)
%\Line(240,0)(220,-15)
%\CCirc(240,0){7}{Black}{Black}
%\Line(335,0)(375,0)
%\Line(280,0)(320,0)
%\CCirc(355,0){5}{Black}{Black}
%\CCirc(300,0){5}{Black}{Black}
%\Oval(420,0)(25,15)(0)
%\CCirc(405,0){7}{Black}{Black}
%\CCirc(420,25){5}{Black}{Black}
%\CCirc(420,-25){5}{Black}{Black}
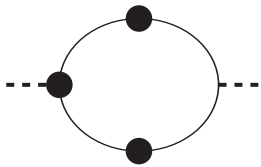
\Text(53,-47)[]{Figure 2: General type of the $\sigma$-field propagator}
\end{picture}
\end{center}
After the $R'$ operation\footnote{The $R'$ operation means that we subtract from the diagram
all divergent subgraphs} we do not have any poles in the integrand for the remaining
one-loop integral. What is left is the finite part containing logarithms of momenta.
 This final integration has the following form:
$$\int\frac{\ln^n(k^2/\mu^2)\ln^m(k^2/p^2)\ln^k(k^2/(k-p)^2)}{k^2(k-p)^2}d^Dk.$$
Due to the naive power counting of divergences in dimensional regularization  we obtain
the result proportional to $\Gamma(2-D/2)$ which is finite for any odd $D$. The
logarithms can not change this property.

To demonstrate how this works explicitly, we consider a particular example of the two-loop
diagram. The result of the $R'$-operation is shown in Fig.3.
\begin{center}
\begin{picture}(300,120)(55,-70)

\Text(-10,1)[]{$R'$} \SetWidth{1.5} \DashLine(0,0)(20,0){3} \DashLine(80,0)(100,0){3}
\DashCArc(50,30)(25,-150,-30){3} \SetWidth{0.5} \Oval(50,0)(30,30)(0) \Text(108,1)[]{=}
\SetWidth{1.5} \DashLine(120,0)(140,0){3} \DashLine(200,0)(220,0){3}
\DashCArc(170,30)(25,-150,-30){3} \SetWidth{0.5} \Oval(170,0)(30,30)(0)
\Text(227,-2)[]{-} \Line(235,0)(295,0) \DashCArc(265,0)(35,0,360){3} \SetWidth{1.5}
\DashCArc(265,0)(20,0,180){3} \DashLine(310,0)(330,0){3} \DashLine(390,0)(410,0){3}
\SetWidth{0.5} \Vertex(360,20){3} \Oval(360,0)(20,30)(0)

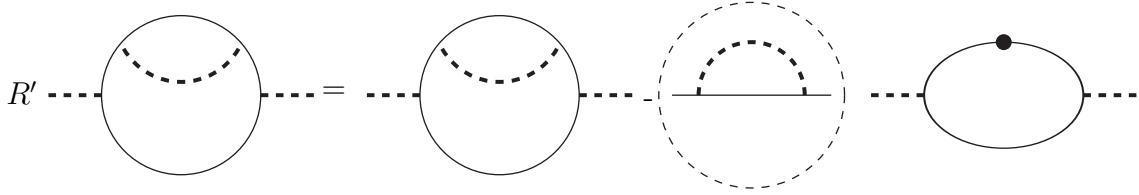
\Text(204,-55)[]{Figure 3: Demonstration of the global divergence cancellation in the two-loop diagram}
\end{picture}
\end{center}
After subtracting the divergence in a subgraph we have prior to the last integration
$$\int\frac{dk}{k^2(p-k)^2}[ \Gamma(-1+\varepsilon)\frac{\Gamma(D/2-1-\varepsilon)
\Gamma(2-\varepsilon)}{\Gamma(D/2-2)\Gamma(D/2+1-2\varepsilon)}\frac{1}{(k^2)^\varepsilon}
+\frac{1}{\varepsilon}\frac{\Gamma(D/2-1)}{\Gamma(D/2-2)\Gamma(D/2+1)} ]\ ,$$ where
$D'=D-2\varepsilon$. The pole terms in the integrand cancel and expanding it over
$\varepsilon$ one gets $\log(k^2)$.  The last integration gives
$$\Gamma(-1+\varepsilon)\frac{\Gamma(D/2-1-\varepsilon)\Gamma(2-\varepsilon)
\Gamma(2+\varepsilon-D'/2)\Gamma(D'/2-1)\Gamma(D'/2-1-\varepsilon)
}{\Gamma(D/2-2)\Gamma(D/2+1-2\varepsilon)\Gamma(1+\varepsilon)
\Gamma(D'-2-\varepsilon)}\frac{p^2}{(p^2)^{2\varepsilon}}$$
$$+\frac{1}{\varepsilon}\frac{\Gamma(D/2-1)}{\Gamma(D/2-2)\Gamma(D/2+1)}
\frac{\Gamma(2-D'/2)\Gamma^2(D'/2-1)}{\Gamma(D'-2)}\frac{p^2}{(p^2)^{\varepsilon}} \ = \
O(1).$$ Thus, after the $R'$ operation the diagram is finite and we do not need the
$\sigma$ field renormalization.

This way one gets the perturbative expansion with only logarithmic divergences. We will
show now that this is not expansion over $\lambda$ with a negative dimension equal
to  $D-4$ but rather $1/N$ expansion with dimensionless coupling.

\section{Properties of the 1/N expansion}

Consider now the leading order calculations. We start with the $1/N$ terms for the propagator
of $\phi$ and the triple vertex. One has the diagrams shown in Fig.4. Notice
that besides the one-loop diagrams in the same order of $1/N$ expansion one has the
two-loop diagram for the vertex.
\begin{center}
\begin{picture}(250,130)(0,-55)
\Line(0,0)(50,0)
\SetWidth{1.5}
\DashCArc(25,0)(15,0,180){3}
\SetWidth{0.5}
\Text(25,-10)[]{a}

\SetWidth{1.5} \DashLine(110,55)(110,35){3} \SetWidth{0.5} \Line(80,-0)(110,35)
\Line(140,-0)(110,35) \SetWidth{1.5} \DashLine(92,10)(128,10){3} \SetWidth{0.5}
\Text(110,-10)[]{b}

\Line(170,0)(250,0)
\SetWidth{1.5}
\DashLine(190,0)(190,25){3}
\DashLine(230,0)(230,25){3}
\SetWidth{0.5}
\Line(230,25)(190,25)
\Line(190,25)(210,45)
\Line(210,45)(230,25)
\SetWidth{1.5}
\DashLine(210,45)(210,60){3}
\SetWidth{0.5}
\Text(210,-10)[]{c}
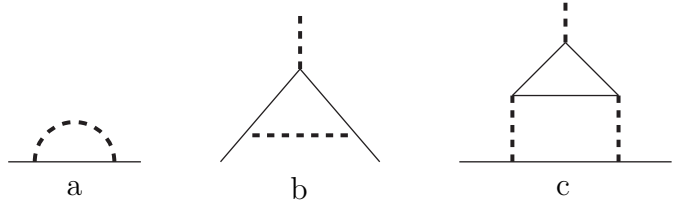
\Text(119,-30)[]{Figure 4: The leading order diagrams giving a contribution to the $\phi$ field propagator}
\Text(3,-43)[]{and the triple vertex  in $1/N$ expansion}
\end{picture}
\end{center}

Let us start with the diagram a). One has
\begin{eqnarray*}
I_a &\sim & \int \frac{d^{D'}k}{(2\pi)^DN}\frac{\lambda}{[(k-p)^2-m^2][1+\lambda f(D)
(-k^2)^{D/2-2}]}\\
&=&\int \frac{d^{D'}k}{(2\pi)^DN}\frac{1}{[(k-p)^2-m^2][1/\lambda+ f(D) (-k^2)^{D/2-2}]},
\ \ D'=D-2\varepsilon .
\end{eqnarray*}
This integral may have problems with evaluation if $f(D)<0$. As we have already mentioned, we
evaluate the integral in a sense of its principle value. Then the UV asymptotics is given by
$$I_a\  \Rightarrow\  \int \frac{d^{D'}k}{(2\pi)^DNf(D)}\frac{1}{(k-p)^2(-k^2)^{D/2-2}}.$$
One can see that the original coupling $\lambda$ plays the role of inverse mass and drops out
from the UV expression. What is left is a dimensionless $1/N$ term.

Calculating the singular parts of the diagrams of Fig.4 in  dimensional
regularization with $D'=D-2\varepsilon$ one finds
\begin{eqnarray}\label{sing}
Diag.a &\Rightarrow& \frac{1}{\varepsilon N}A,\ \ \ Diag.b\ \Rightarrow\
\frac{1}{\varepsilon N}B,\ \ \
Diag.c \ \Rightarrow\ \frac{1}{\varepsilon N}C,\\
&& \nonumber \\ &&\hspace*{-2.5cm}
A=\frac{2\Gamma(D-2)}{\Gamma(D/2-2)\Gamma(D/2-1)\Gamma(D/2+1)\Gamma(2-D/2)}, \ \ \
B=\frac{D}{4-D}A, \ \ \ C=\frac{D(D-3)}{4-D}A. \nonumber
\end{eqnarray}
The corresponding renormalization constants in the $\overline{MS}$ scheme are then
\begin{eqnarray}\label{z}
Z_2^{-1}&=&1-\frac{1}{\varepsilon}\frac{A}{N},\\
Z_1&=&1-\frac{1}{\varepsilon}\frac{B+C}{N}.
\end{eqnarray}
There is no any coupling in these formulas except for $1/N$. However, when renormalizing
the coupling, i.e., replacing the bare coupling with the renormalized one times the
corresponding Z factors, we face the problem: one cannot take a bare (infinite) number
of components and renormalize it. To overcome this difficulty, we introduce a new
dimensionless coupling $h$ associated with the triple vertex (and not with the $\sigma$
propagator) as
$$ {\cal L}_{int} = -\frac{\sqrt{h}\sqrt{\lambda}}{2\sqrt{N}}\sigma \vec{\phi}^2.$$
Then in the leading order in $1/N$ the renormalization constants and the coupling take
the form
\begin{eqnarray}\label{r}
Z_2^{-1}&=&1-\frac{h}{\varepsilon}\frac{A}{N},\\
Z_1&=&1-\frac{h}{\varepsilon}\frac{B}{N}-\frac{h^2}{\varepsilon}\frac{C}{N}, \\
h_B &=& (\mu^2)^\varepsilon h
Z_1^2Z_2^{-2}=h\left(1-\frac{h}{\varepsilon}\frac{2(A+B)}{N}-
\frac{h^2}{\varepsilon}\frac{C}{N}\right).
\end{eqnarray}

This is not, however, the final expression. To see this, we consider the next order of
$1/N$ expansion. The corresponding diagrams for the $\phi$ propagator are shown in Fig.5.
Again one can see that the $1/N^2$ terms contain not only the two-loop diagrams but also
the three- and even four-loop ones.

\begin{center}
\begin{picture}(400,235)(0,-170)

\Line(0,0)(80,0)
\SetWidth{1.5}
\DashCArc(40,0)(20,0,180){3}
\DashCArc(40,0)(30,0,180){3}
\SetWidth{0.5}
\Line(100,0)(180,0)
\Text(40,-30)[]{a}
\SetWidth{1.5}
\DashCArc(130,0)(20,0,180){3}
\DashCArc(150,0)(20,180,360){3}
\SetWidth{0.5}
\Text(140,-30)[]{b}
\Line(200,0)(280,0)
\SetWidth{1.5}
\DashCArc(220,0)(10,0,180){3}
\DashCArc(260,0)(10,0,180){3}
\SetWidth{0.5}
\Line(300,0)(380,0)
\Text(240,-30)[]{c}
\SetWidth{1.5}
\DashLine(320,0)(320,45){3}
\DashLine(340,0)(340,30){3}
\DashLine(360,0)(360,45){3}
\SetWidth{0.5}
\Line(320,45)(360,45)
\Line(320,45)(340,30)
\Line(340,30)(360,45)
\Text(340,-30)[]{d}
\Line(0,-100)(100,-100)
\Oval(50,-65)(15,20)(0)
\Text(50,-120)[]{e}
\SetWidth{1.5}
\DashCArc(50,-100)(40,120,180){3}
\DashCArc(50,-100)(40,0,60){3}
\DashCArc(50,-45)(15,-140,-40){3}
\SetWidth{0.5}
\Line(120,-100)(220,-100)
\Oval(170,-65)(15,20)(0)
\SetWidth{1.5}
\DashCArc(170,-100)(40,120,180){3}
\DashCArc(170,-100)(40,0,60){3}
\DashLine(170,-50)(170,-80){3}
\SetWidth{0.5}
\Line(250,-100)(350,-100)
\Text(170,-120)[]{f}
\SetWidth{1.5}
\DashCArc(280,-100)(20,90,180){3}
\DashCArc(320,-100)(20,0,90){3}
\SetWidth{0.5}
\Line(280,-80)(290,-70)
\Line(280,-80)(290,-90)
\Line(290,-70)(290,-90)
\SetWidth{1.5}
\DashLine(290,-70)(310,-70){3}
\DashLine(290,-90)(310,-90){3}
\SetWidth{0.5}
\Line(310,-70)(320,-80)
\Line(310,-90)(320,-80)
\Line(310,-70)(310,-90)
\Text(300,-120)[]{g}

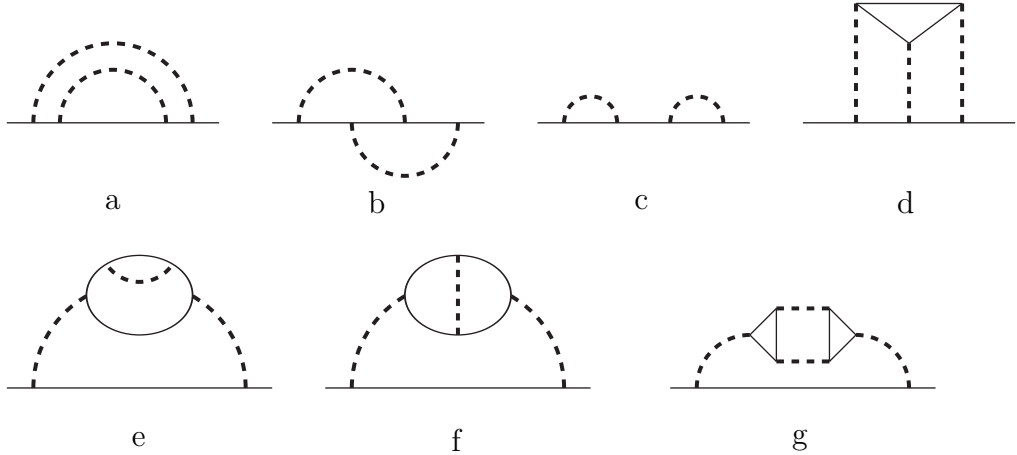
\Text(192,-140)[]{Figure 5: The second order diagrams giving a contribution
 to the $\phi$ field propagator}
\Text(22,-152)[]{in $1/N$ expansion}
\end{picture}
\end{center}

All these diagrams are double logarithmically divergent, i.e., contain both single and
double poles in dimensional regularization. We calculate the leading double pole after
subtraction of the divergent subgraphs, i.e. perform  the $R'$-operation. The answer is:
\begin{eqnarray}\label{sing2}
Diag.a &\Rightarrow& -\frac{1}{\varepsilon^2 N^2}\frac 12A^2h^2,\ \ \ Diag.b\
\Rightarrow\ -\frac{1}{\varepsilon^2 N^2}ABh^2,\ \ \
Diag.c \ \Rightarrow\ -\frac{1}{\varepsilon^2 N^2}A^2h^2, \nonumber \\
Diag.d &\Rightarrow& -\frac{1}{\varepsilon^2 N^2}\frac 43 ACh^3,\
\ \ Diag.e\ \Rightarrow\ \frac{1}{\varepsilon^2 N^2}\frac 23
A^2h^3,\ \ \ Diag.f \ \Rightarrow\ \frac{1}{\varepsilon^2
N^2}\frac 23 ABh^3,\nonumber \\ Diag.g & \Rightarrow&
\frac{1}{\varepsilon^2 N^2}ACh^4.
\end{eqnarray}
We performed the same calculation for the vertex diagrams, but there are too many of them
to reproduce, so we present below only the final answer for the Z-factors.

 Here we face a problem, namely,
in subtracting the divergent subgraphs in the graphs e-g, we get the diagram which is
absent in our expansion, since it is already included in our bold $\sigma$ line (see
Fig.6).
\begin{center}
\begin{picture}(110,95)(0,-35)
\Line(0,0)(100,0) \Oval(50,35)(15,20)(0) \SetWidth{1.5} \DashCArc(50,0)(40,120,180){3}
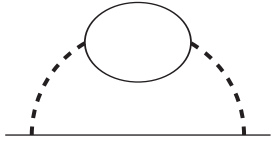
\DashCArc(50,0)(40,0,60){3} \SetWidth{0.5} \Text(50,-20)[]{Figure 6: The "forbidden" loop
diagram}
\end{picture}
\end{center}
There would be no problem unless this diagram is needed to match the so-called pole
equations~\cite{Hooft} which allow one to calculate the higher order poles in the Z factors
from the single one. However, if we include this diagram in the $\sigma$ line, it will
not change the latter, except for the additional $h$ factor coming from the vertex and not
compensated by the propagator. Apparently, one can continue this insertion procedure and
add any number of such loops not changing the order of $1/N$ expansion. The result is the
sum of a geometrical progression
\begin{eqnarray*}
&&\frac{1}{1/\lambda-{\cal O}}+\frac{1}{1/\lambda-{\cal O}}h{\cal O}\frac{1}{1/\lambda-{\cal
O}}+\frac{1}{1/\lambda-{\cal O}}h{\cal O}\frac{1}{1/\lambda-{\cal
O}}h{\cal O}\frac{1}{1/\lambda-{\cal
O}} +... \\
&=& \frac{1}{1/\lambda-{\cal O}}\left( \frac{1}{1-h{\cal O}/(1/\lambda-{\cal O})}\right)=
\frac{1}{1/\lambda-(1+h){\cal O}}.
\end{eqnarray*}
 Altogether this leads to the following effective Lagrangian for  $1/N$ perturbation theory
\begin{equation}\label{ef}
 {\cal L}_{eff} = \frac 12 (\partial_\mu \vec{\phi})^2-
 \frac{\sqrt{h}}{2\sqrt{N}}\sigma(\vec{\phi}^2)+\frac{1}{2\lambda} \sigma^2
 +\frac 12 f(D)\sigma (\partial^2)^{D/2-2}\sigma (1+h).
\end{equation}
This means that in UV regime one should multiply every $\sigma$ line by $1/(1+h)$.

Having all this in mind we come to the final expressions for the Z factors within the
$1/N$ expansion:
\begin{eqnarray}\label{new}
Z_1&=&1-\frac{1}{\varepsilon N}\left(\frac{B
h}{1+h}+\frac{Ch^2}{(1+h)^2}\right)+\frac{1}{\varepsilon^2 N^2}\left( \frac
32\frac{B^2h^2}{(1+h)^2} + \frac{ABh^2}{(1+h)^2}\right.\nonumber \\ &&\left. + \frac
{11}{3}\frac{BCh^3}{(1+h)^3} + \frac 43\frac{ACh^3}{(1+h)^3} - \frac
23\frac{ABh^3}{(1+h)^3} - \frac 23\frac{B^2h^3}{(1+h)^3} - 2\frac{BCh^4}{(1+h)^4}
\right.\nonumber \\ &&\left. - \frac{ACh^4}{(1+h)^4} + \frac 52\frac{C^2h^4}{(1+h)^4} -
\frac 85 \frac{C^2h^5}{(1+h)^5}\right)
+O(\frac{1}{\varepsilon N^2}), \\
Z_2^{-1}&=&1-\frac{1}{\varepsilon N}\frac{Ah}{1+h}+\frac{1}{\varepsilon^2 N^2}\left(
\frac 32\frac{A^2h^2}{(1+h)^2}+\frac{ABh^2}{(1+h)^2}-\frac
23\frac{A^2h^3}{(1+h)^3}\right.\nonumber \\ &&\left. - \frac 23\frac{ABh^3}{(1+h)^3}
+\frac 43\frac{ACh^3}{(1+h)^3}-\frac{ACh^4}{(1+h)^4}\right)+O(\frac{1}{\varepsilon N^2}).
\label{new2}
\end{eqnarray}

In addition on has also the renormalization of $1/\lambda$ parameter which plays the role
of a mass of the $\sigma$ field. In the leading order of $1/N$ expansion the relevant
diagrams are shown in Fig.7.
\begin{center}
\begin{picture}(400,120)(0,-160)

\Oval(50,-65)(15,20)(0) \Text(50,-90)[]{a} \SetWidth{1.5} \DashLine(10,-65)(30,-65){3}
\DashLine(70,-65)(90,-65){3} \DashCArc(50,-45)(15,-140,-40){3} \SetWidth{0.5}
\Line(46,-56)(54,-64) \Line(54,-56)(46,-64) \Oval(170,-65)(15,20)(0) \SetWidth{1.5}
\DashLine(130,-65)(150,-65){3} \DashLine(190,-65)(210,-65){3}
\DashLine(170,-50)(170,-80){3} \SetWidth{0.5} \Line(166,-61)(174,-69)
\Line(174,-61)(166,-69) \Text(170,-90)[]{b} \SetWidth{1.5} \DashLine(260,-65)(280,-65){3}
\DashLine(320,-65)(340,-65){3} \SetWidth{0.5} \Line(280,-65)(290,-55)
\Line(280,-65)(290,-75) \Line(290,-55)(290,-75) \SetWidth{1.5}
\DashLine(290,-55)(310,-55){3} \DashLine(290,-75)(310,-75){3} \SetWidth{0.5}
\Line(296,-51)(304,-59) \Line(304,-51)(296,-59) \Line(310,-55)(320,-65)
\Line(310,-75)(320,-65) \Line(310,-55)(310,-75) \Text(300,-90)[]{c}

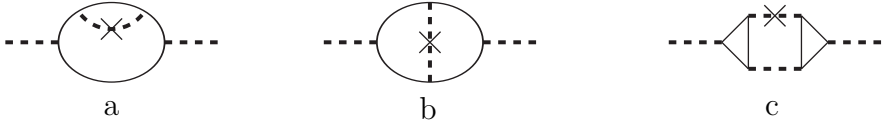
\Text(192,-110)[]{Figure 7: The leading order diagrams giving a contribution
 to  $1/\lambda$ renormalization}
\end{picture}
\end{center}\vspace{-1cm}
Taking into account the previous discussion we obtain the following results
\begin{eqnarray}\label{lambda}
Diag.a &\Rightarrow& \frac{h^2}{\varepsilon N (1+h)^2}F,\ \ Diag.b\ \Rightarrow\
\frac{h^2}{\varepsilon N (1+h)^2}E,\ \
Diag.c \ \Rightarrow\ \frac{h^3}{\varepsilon N  (1+h)^3 }G,   \nonumber \\
&& \nonumber \\ &&\hspace*{-2.5cm} F=\frac{2 (D-3)(D-6) \Gamma(D-2) }{\Gamma(D/2)
\Gamma^2(D/2-1) \Gamma(2-D/2)}, \ \ \ E=\frac{D-4}{D-6}F, \ \ \ G=\frac{4 (D-3) }{3 (D-6)
}F.
\end{eqnarray}
which gives the renormalization constant for $1/\lambda$
\begin{equation}
Z_{1/ \lambda} \ = \ 1 - \frac{1}{N \varepsilon} \left( \frac{(F+E)h^2}{(1+h)^2}
 + \frac{Gh^3}{(1+h)^3} \right).
\end{equation}

\section{Renormalization group in 1/N expansion}

Having these expressions for the Z factors one can construct the coupling constant
renormalization and the  corresponding RG functions. One has as usual in the dimensional
regularization
\begin{eqnarray}\label{rg}
  h_B&=&(\mu^2)^\varepsilon hZ_1^2Z_2^{-2}=(\mu^2)^\varepsilon \left(
  h+\sum_{n=1}^{\infty}\frac{a_n(h,N)}{\varepsilon^n}\right),\\
  Z_i&=&1+\sum_{n=1}^{\infty}\frac{c^i_n(h,N)}{\varepsilon^n},\label{zz}
\end{eqnarray}
where the first coefficients $a_n$ and $c^i_n$ can be deduced from
eqs.(\ref{new},\ref{new2}).

This allows one to get the anomalous dimensions and the beta function defined as
\begin{eqnarray}\label{anom}
  \gamma(h,N)&=&-\mu^2\frac{d}{d\mu^2}\log Z = h\frac{d}{dh}c_1, \\
 \beta(h,N)&=&2h(\gamma_1+\gamma_2)=(h\frac{d}{dh}-1)a_1.
\end{eqnarray}
With the help of eqs.(\ref{new},\ref{new2}) one gets in the leading order of $1/N$
expansion\footnote{Note that the anomalous dimension of a field $\gamma_2$, is defined
with respect to $Z_2^{-1}$.}
\begin{eqnarray}\label{dim}
  \gamma_2(h,N)&=&-\frac 1N \frac{Ah}{(1+h)^2}, \ \ \
  \gamma_1(h,N)=-\frac 1N \left(\frac{Bh}{(1+h)^2}+\frac{2Ch^2}{(1+h)^3}\right),\\
  \beta(h,N)&=&-\frac 1N \left(\frac{2(A+B)h^2}{(1+h)^2}+\frac{4Ch^3}{(1+h)^3}\right).
\end{eqnarray}

It is instructive to check the so-called pole equations~\cite{Hooft} that express the
coefficients of the higher order poles in $\varepsilon$ of the Z factors via the coefficients
of a simple pole. For $Z_2^{-1}$ one has, according to (\ref{new2}),
\begin{eqnarray}\label{poles}
  c_1(h,N)&=&-\frac 1N \frac{Ah}{1+h}, \\
  c_2(h,N)&=& \frac{1}{N^2}
  \left(
\frac 32\frac{A^2h^2}{(1+h)^2}+\frac{ABh^2}{(1+h)^2}-\frac
23\frac{A^2h^3}{(1+h)^3}\right.\nonumber \\ &&\left. - \frac 23\frac{ABh^3}{(1+h)^3}
+\frac 43\frac{ACh^3}{(1+h)^3}-\frac{ACh^4}{(1+h)^4}\right).\label{poles2}
\end{eqnarray}
At the same time the coefficient $c_2$ can be expressed through $c_1$ via the pole equations
as
\begin{equation}\label{p}
  h\frac{dc_2}{dh}=\gamma_2c_1+\beta\frac{dc_1}{dh},
\end{equation}
which gives
$$h\frac{dc_2}{dh}=\frac{1}{N^2}\frac{Ah}{(1+h)^2}\frac{Ah}{1+h}
+\frac{1}{N^2}\left(\frac{2(A+B)h^2}{(1+h)^2}+\frac{4Ch^3}{(1+h)^3}\right)\frac{A}{(1+h)^2}.$$
Integrating this equation one gets for $c_2$ the expression coinciding with
(\ref{poles2}) which was obtained by direct diagram evaluation. Notice that to get this
coincidence the $h$-dependence in the denominator of eqs.(\ref{new},\ref{new2}) was
absolutely crucial.

We have also checked  the pole equations for the renormalized coupling. Combining
eqs.(\ref{new}) and (\ref{new2}) one gets for thre renormalization constant
$Z_h=Z_1^2Z_2^{-2}$
\begin{eqnarray}\label{poles}
  a_1(h,N)&=&-\frac 1N \left( \frac{2(A+B)h^2}{1+h} + \frac{4Ch^3}{(1+h)^2}\right), \\
  a_2(h,N)&=& \frac{1}{N^2}
  \left(
4\frac{(A+B)^2h^3}{(1+h)^2} - \frac43\frac{(A+B)^2h^4}{(1+h)^3}+ \frac
{28}{3}\frac{(A+B)Ch^4}{(1+h)^3} \right.\nonumber \\ &&\left. -4\frac{(A+B)Ch^5}{(1+h)^4}
+ 6\frac{C^2h^5}{(1+h)^4} - \frac{16}{5}\frac{C^2h^6}{(1+h)^5}\right).\label{poles3}
\end{eqnarray}
At the same time from the pole equations one has
\begin{equation}\label{rg}
(h\frac{d}{dh}-1)a_n=\beta\frac{da_{n-1}}{dh}.
\end{equation}
We checked that the coefficient $a_2$ evaluated this way coincides with (\ref{poles3})
which was obtained by direct diagram evaluation.

 At last, in the leading order in $h$ when
$$a_1(h,N)\simeq-\frac{2(A+B)h^2}{N}\ \ \mbox{and} \ \  \beta(h,N)\simeq-\frac{2(A+B)h^2}{N}$$
eq.(\ref{rg}) gives a geometrical progression
$$a_n(h,N)=a_1(h,N)^n.$$
We have checked this relation up to third order in $1/N$ expansion
and confirmed its validity.

Having expression for the $\beta$ function one may wonder how the coupling is running.
The crucial point here is the sign of the $\beta$ function. One has
\begin{equation}\label{beta}
\frac{dh}{dt}=\beta(h,N)=-\frac 1N \frac{\displaystyle
4\Gamma(D-2)\left(\frac{2h^2}{(1+h)^2}+\frac{D(D-3)h^3}{(1+h)^3}
\right)}{\Gamma(D/2-2)\Gamma(D/2-1)\Gamma(D/2+1)\Gamma(3-D/2)}.
\end{equation}
The beta function can also be rewritten as
\begin{equation}\label{beta2}
\beta(h,N)=-\frac 1N \frac{\displaystyle
2^{D-1}\Gamma(D/2-1/2)(-)^{(D-1)/2}\left(\frac{2h^2}{(1+h)^2}+\frac{D(D-3)h^3}{(1+h)^3}
\right)}{\Gamma(1/2)\pi\Gamma(D/2+1)},
\end{equation}
that clearly indicates the the theory is UV asymptotically free for $D=2k+1,\ k$ - even
and IR free for $k$-odd. Solution of the RG equation looks somewhat complicated, but for
the small coupling in the leading order it simply equals the usual leading log
approximation ($t=\log(Q^2/\mu^2)$
\begin{equation}\label{sol}
  h(t,h) \simeq \frac{h}{1-\beta_0ht}, \ \ \ \beta_0=-\frac 1N \frac{\displaystyle
2^{D}\Gamma(D/2-1/2)(-)^{(D-1)/2}}{\Gamma(1/2)\pi\Gamma(D/2+1)}.
\end{equation}
For example, for $D=5,7$ the beta function equals $\beta_0=-256/15\pi^2N$ and
$2^{12}/105\pi^2N$, respectively.

\section{Conclusion}

We conclude that even in a formally non-renormalizable theory it is possible to construct
renormalizable $1/N$ expansion which obeys all the rules of a usual perturbation theory.
The expansion parameter is dimensionless, the coupling is running logarithmically, all
divergencies are absorbed into the renormalization of the wave function and the coupling.
The propagator of the spurion field $\sigma$ behaves as $1/(p^2)^{D/2-2}$ which provides
a better convergence of the diagrams. In the Euclidean space, depending on the value of
$D$ it either contains no additional singularities, or has a simple renormalon type pole.
The original dimensionful coupling plays a role of a mass and is renormalized
multiplicatively. Expansion over this coupling is singular and creates the usual
nonrenormalizable terms.

Properties of $1/N$ expansion do not depend on the space-time dimension if it is odd. For
an even dimension our formulas have a singularity which originates from divergence of
simple one-loop bubbles that are summed up in the denominator of the $\sigma$ field
propagator. This requires a special treatment.

In fact, the $1/N$ expansion was considered in the same way in the 3 dimensional nonlinear
$\sigma$-model which is also non-renormalizable in three dimensions~\cite{Aref3}. It was shown that
the resulting PT exhibits the properties of a renormalizable theory. Here we went
further, we calculated the leading divergencies, checked the RG properties of the new
expansion and revealed its nonpolynomial character.

There is one essential point that we omitted in our discussion. This is the analytical
properties of the $\sigma$ field propagator and the unitarity of a resulting theory. As
one can see from eq.(\ref{ppp}) the propagator of the $\sigma$ field contains a cut
starting from $p^2=0$ in massless case, thus leading to K\"allen-Lehman  representation
with continuous spectral function. This can be interpreted as continuous spectrum of
states. Then, the theory happens to be unitary in the full space including these states.
We postpone  discussion of this problem to a separate publication.

\section*{Acknowledgements}
Financial support from RFBR grant \# 05-02-17603 and grant of the Ministry of Education and Science
of the Russian Federation \# 2339.2003.2 is kindly acknowledged. G.V.
would like to thank Dynasty Foundation for support. We are grateful to I.Aref'eva and
S.Mikhailov for valuable discussions.

\end{document}